# Heteronuclear and Homonuclear Vector Solitons in Lasers

Yueqing Du,[1] Zhenzhu Zhang,[1] Heze Zhang,[1] Jia Xue,[1] Chao Zeng,[1, 2] Yudong Cui,[3] Dong Mao,[1, *] Boris A. Malomed,[4, 5] and Jianlin Zhao[1, †]

[1]*Shaanxi Key Laboratory of Optical Information Technology, Key Laboratory of Light Field Manipulation and Information Acquisition, Ministry of Industry and Information Technology, School of Physical Science and Technology, Northwestern Polytechnical University, Xi'an 710129, China*
[2]*Shenzhen Research Institute of Northwestern Polytechnical University, Shenzhen 518057, China*
[3]*State Key Laboratory of Extreme Photonics and Instrumentation, College of Optical Science and Engineering, Zhejiang University, Hangzhou, 310027, China*
[4]*Department of Physical Electronics, Faculty of Engineering, and Center for Light-Matter Interaction, Tel Aviv University, P.O.B. 39040, Ramat Aviv, Tel Aviv, Israel*
[5]*Instituto de Alta Investigación, Universidad de Tarapacá, Casilla 7D, Arica 1000000, Chile*

Vector solitons (VSs), being observed across various fields from optics to Bose-Einstein condensates, are localized structures composed of orthogonal modes bound by nonlinear couplings. Nevertheless, the influence of intermodal linear coupling on the physical properties of this bimodal structure remains to be decently revealed and harnessed. Utilizing an ultrafast fiber laser as a platform, we predict and demonstrate that the linear mode coupling (LMC) induces the deformable VS in terms of the temporal and spectral structures. Weak LMC supports heteronuclear vector solitons built of dissimilar polarization modes, i.e., a single pulse coupled to an orthogonal damped pulse chain. On the other hand, strong LMC facilitates the homonuclear VS composed of polarization modes with similar structures, in the form of soliton compounds featuring "caterpillar" motions. Our findings reveal new patterns of VSs and open an effective avenue for versatile ultrafast optical sources.

Vector solitons (VSs), i.e., localized structures composed of bound orthogonal modes, are ubiquitous patterns in various physical systems, including optics [1-4], elastic metamaterials [5,6], and Bose-Einstein condensates [7,8]. The VS offers a versatile testbed for studying complex physics and soliton interactions. Diverse phenomena associated with VSs have been explored in optical systems, e.g., collisions between VSs [9-13] and emergent composite structures of VSs [14-18]. The stability of VS relies on the nonlinear mode coupling, which is mostly mediated by the cross-phase modulation (XPM) [19-25]. In dissipative systems, such as lasers or driven resonators, temporal VSs are sustained by the interplay of gain, loss, dispersion, nonlinearity, and intermodal interactions [25-38]. Hence, the laser is a compact platform for exploring uncharted VS phenomena [39] and studying interdisciplinary topics such as vectorial rogue waves [3] and polarization synchronization of bimodal wavepackets [40].

In lasers, initial noises evolve into VSs through self-organization, which is sensitive to XPM and birefringence. Naturally, the adjustment of the pump power and polarization controllers (PCs) [25] is utilized to manipulate VS properties. Nevertheless, well-defined mappings between control parameters and soliton properties, e.g., the temporal and spectral structures, are lacking. Moreover, XPM facilitates similar pulse shapes and spectral profiles for polarization modes of temporal VSs [32-34] that can be categorized as the homonuclear one, following the analogy with the diatomic molecule [41]. In this regard, the diversity and controllability of VSs are limited when constituent modes are coupled solely through XPM.

On the other hand, as a vital effect in broad scientific fields such as cold atomic gases [42] and quantum circuits [43], the linear mode coupling (LMC) [44-48] is also acknowledged to be significant for optics in coupled cavities [49,50], spatiotemporal solitons [51-53] and VSs [54]. The ability of LMC to manipulate vector and spatiotemporal solitons relies on intermodal mixing, which modifies the pulse profile and intermodal energy distribution [51,54]. Besides, solitons are localized structures that can be temporally shifted to generate addressable patterns [2,55-58]. Consequently, the relative time shift between polarization modes within the VS has a fundamental influence on the coupling scenario.

Thus, the interplay of LMC with the time shift may pave an avenue to manipulate and synthesize unprecedented structures of VSs. Such a dual-manipulation concept bears a certain resemblance to the polarization-time entanglement of quantum walks [59]. Note that the time shift can be applied either to the entire VS or to its particular polarization mode. We term the latter option as "intramodal time shift" (ITS).

In this work, we demonstrate a class of VSs with deformable temporal and spectral structures leveraging LMC and ITS in the fiber laser. Increasing the LMC strength promotes a transition from a VS built of dissimilar modes (categorized as heteronuclear VSs, as suggested by the similarity to heteronuclear soliton molecule [60]), to a homonuclear VS. The emergent spectral and temporal properties, such as the tunable damped pulse chain with a comb-like spectrum, and soliton compounds performing the caterpillar-like motion, are all inaccessible species for VSs without LMC. Our work demonstrates that the LMC-ITS interplay can effectively manipulate the VS characteristics, which can be extended to research soliton-based applications [61-65], nonlinear non-Hermitian photonics [66], and physics in photonic lattices [67].

We address the laser configuration displayed in Fig. 1(a). In the resonator, LMC between the vertical and horizontal modes is introduced by a fiber PC [54]. To implement ITS, the polarization modes are separated into two branches by a polarization beam splitter, with a delay between them adjusted by a tunable delay line. After the reconnection of the branches, the dislocated modes are orthogonally recombined by the polarization beam combiner, see details in Sec. 1.1 of Supplemental Material (SM) [68].

As the two modes evolve synchronously after the VS forms, the relative delay between the recombined polarization modes is determined by the time shift of the horizontal mode, i.e., ITS, while the vertical mode remains fixed, as schematically shown in Fig. 1(b). In experiments, the ITS spans from -187 to 213 ps (see Sec. 1.2 in SM [68]). Note that configurations similar to the present ITS setup have been used to realize the quantum walk [59] and non-Abelian gauge fields [67]. Overall, the LMC+ITS interplay will be shown below to exert significant effects on VSs.

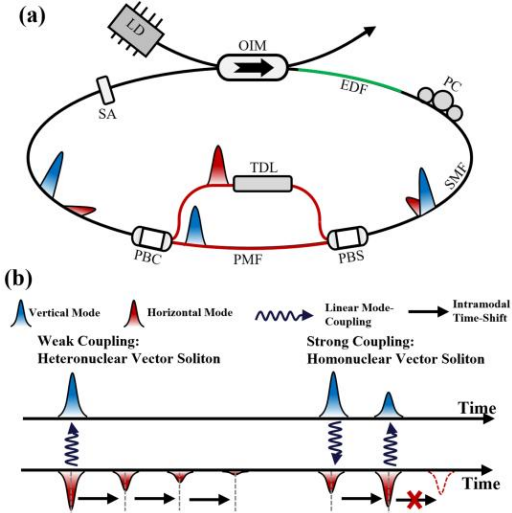

FIG. 1. (a) The fiber-laser setup. (b) The illustration of the deformable VS governed by the interplay of LMC and ITS. LD: laser diode. EDF: Erbium-doped fiber. SMF: single-mode fiber. OIM: Optical integrated module. PBS: polarization beam splitter. TDL: tunable-delay line. PBC: polarization beam combiner. PMF: polarization-maintaining fiber. PC: polarization controller. SA: saturable absorber.

*Simulations.*—To explore the impact of LMC on VSs, we first conduct numerical simulations of the ring-cavity model. In erbium-doped and single-mode fibers, the VS is described by the coupled generalized nonlinear Schrödinger equations, while the modes separated by PBS are individually governed by the nonlinear Schrödinger equations. Details of the simulations are given in Sec. 1.3 of SM [68].

As mentioned above, ITS is expressed as a time shift of the horizontal mode: $v(z, t) \to v(z, t - t_s)$, wherein $t_s$ is the ITS value. The PC-induced LMC between the two modes is defined as [54]:

$$\begin{bmatrix} u_o \\ v_o \end{bmatrix} = \begin{bmatrix} t_u & \kappa \\ \kappa & t_v \end{bmatrix} \times \begin{bmatrix} u_i \\ v_i \end{bmatrix}, \quad (1)$$

where $u_o$ ($u_i$) and $v_o$ ($v_i$) are the amplitudes of the vertical and horizontal modes after (before) the LMC, respectively. Further, $\kappa = \sin(2\theta)(e^{i\varphi} - 1)/2$ is the LMC coefficient, where $t_v = (e^{i\varphi} - 1)\cos^2\theta + 1$ and $t_u = (e^{i\varphi} - 1)\sin^2\theta + 1$ are the self-transmission coefficients of the horizontal and vertical modes, respectively. The coupling strength $|\kappa|^2 = \sin^2(2\theta)\sin^2(\varphi/2)$ can be tuned by varying the rotation angle $\theta$ and phase retard $\varphi$ of PC. The fiber PC can prevent disturbances caused by environmental fluctuations, compared to other types of PCs composed of rotatable waveplates.

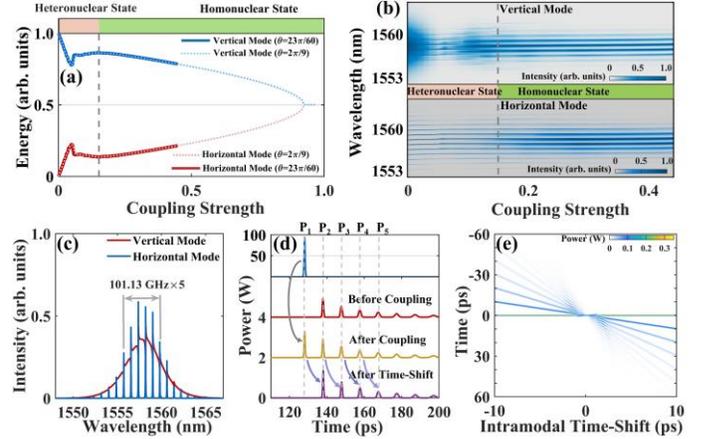

FIG. 2. Simulation results of state transition of the VS versus the coupling strength, produced by varying phase retard $\varphi$ with fixed angle $\theta$. The gain-saturation energy $E_s$ and ITS are 20 pJ and 10 ps, respectively. (a) The normalized energies of the two modes versus the coupling strength for $\theta = 23\pi/60$ and $2\pi/9$ (solid and dotted curves, respectively). (b) The spectral profiles of the vertical and horizontal modes (top and bottom panels, respectively) versus the coupling strength for $\theta = 23\pi/60$. The "heteronuclear" VS with weak LMC, corresponding to $|\kappa|^2 = 0.014$ ($\theta = 23\pi/60$, $\varphi = \pi/9$): (c) spectra of the two modes. (d) Profiles of the vertical mode (the top panel) and horizontal one at three cavity positions (three bottom panels). The arrows plotted in (d) are defined in the main text. (e) Profiles of the horizontal mode at the laser output versus values of ITSs.

We investigate the response of the VS and its modes by increasing the LMC strength from 0 to 0.45 (e.g., $\theta = 23\pi/60$, $\varphi$ changes from 0 to $\pi$), meanwhile, ITS is chosen to be much larger than the pulse duration (~350 fs), e.g., 10 ps. The variation of the energies [Fig. 2(a)] and spectra [Fig. 2(b)] of the two modes reveals a transition point at a critical LMC strength, $|\kappa|^2 = 0.17$. Below and above this point, the VS features the heteronuclear and homonuclear states, respectively. Specifically, the heteronuclear VS is composed of distinct modes, one of which features a broadband spectrum, while the other has a discrete comb-like spectrum. Oppositely, the two modes of the homonuclear VS exhibit similar modulated spectral profiles [Fig. 2(b)]. The deformation from heteronuclear soliton to homonuclear soliton can be elucidated as follows: the enhanced LMC induces stronger intermodal energy exchange and modes mixing, which suppresses the intermodal difference in the energy, profile, and spectrum. This process is essentially a soliton phase transition analogous to the one tunable by mode coupling in non-Hermitian photonics [73]. Although $\theta$ determines the range of LMC strength [from 0 to $\sin^2(2\theta)$], the conclusion that the VS transforms from the heteronuclear to homonuclear state with the increased LMC strength is valid for all values of $\theta$. We have also verified that the formed VSs are independent of initial conditions, indicating their strong attractor properties.

In the heteronuclear state [Fig. 2(c)], the horizontal mode has multiple coherent combs aligned with spectral dips of the vertical mode, indicating the LMC-induced intermodal energy coupling. The 101.13-GHz comb spacing is approximately equal to the inverse of 10-ps ITS. In the temporal domain, the vertical mode is a single pulse with a fixed temporal position [the upper panel in Fig. 2(d)], while the horizontal mode is represented by the damped chain of subpulses equally spaced by 9.8-ps [the second panel in Fig. 2(d)], with five strongest subpulses marked as $P_1$-$P_5$. The profiles of the two modes reveal that the vertical single pulse excites the highest subpulse $P_1$ in the horizontal one via LMC (the gray arrow). Then, in the horizontal mode, $P_n$ is coupled to $P_{n+1}$ ($n$ = 1, 2, 3, 4) by ITS, as marked by blue arrows in Fig. 2(d), with gradually decaying amplitudes. As the subpulses are pinned to fixed temporal sites by ITS, the temporal structure of the horizontal mode can be tuned by ITS [Fig. 2(e)].

Next, we elucidate the self-consistent mechanism supporting the damped pulse chain. Due to ITS, each subpulse $P_n$ is maintained by the coupling with the adjacent stronger and narrower subpulse $P_{n-1}$, counteracting the cavity loss and dispersion-induced broadening, thus enabling the self-consistent intracavity evolution of the heteronuclear VS, see further details in Sec. 2.1 of SM [62].

With the LMC strength increasing beyond the critical value of 0.17, the individual subpulses in the horizontal mode become more pronounced, thereby transforming the vertical one into a multi-subpulse soliton compound. Eventually, the two modes assemble a homonuclear VS with a modulated spectrum, e.g., the one shown in Fig. 3(a) when the LMC strength is $|\kappa|^2$ = 0.44. The temporal profiles of both modes feature three unequal subpulses [Fig. 3(b)] with a temporal spacing close to ITS. It is confirmed by the experiments and simulations that the frequency spacing of spectral fringes is inversely proportional to ITS, and the subpulse spacing is close to ITS. The homonuclear VS is different from the soliton molecule in which several permanent subpulses are bound together [4,60,70,71]. Here, the homonuclear VS is built of subpulses that collectively advance in the temporal domain by rolling forward in the direction defined by ITS [Fig. 3(c)], which resembles the caterpillar motion. The inherent periodicity of the caterpillar motion leads to the periodic energy and spectral evolutions [Fig. 3(d)].

To understand the rolling-forward ("caterpillar") regime, we present results of the intracavity evolution over an entire caterpillar period of five roundtrips [Figs. 3(e) and 3(f)]. In the second roundtrip, the vertical $P_1$ transfers a part of its energy to its horizontal counterpart via LMC (the lower gray arrow); then, in the horizontal mode, the so-reinforced $P_1$ boosts $P_2$ via ITS (the short blue arrow); subsequently, in the third roundtrip, the boosted horizontal $P_2$ feeds its vertical counterpart $P_2$ through LMC (the upper gray arrow). After one caterpillar period, the combined effect of the strong LMC and ITS leads to the decay of $P_1$ and enhancement of $P_2$ in the vertical mode, i.e., the *persistent displacement* of the dominant subpulse, which is the mechanism underlying the caterpillar motion.

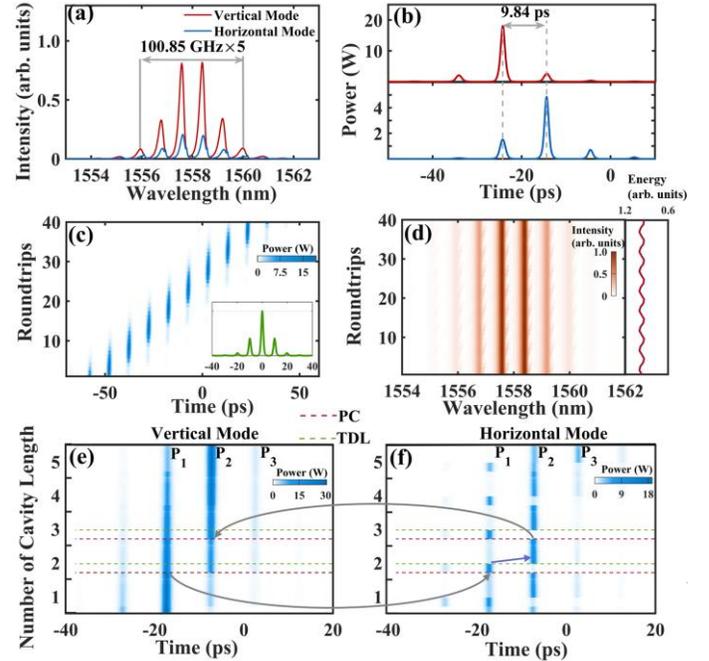

FIG. 3. Simulation results for the homonuclear VS, built of mutually similar vertical and horizontal modes, by increasing LMC strength to 0.44 ($\theta$ = 23$\pi$/60, $\varphi$ = 17$\pi$/18, $E_s$ = 20 pJ, ITS = 10 ps). (a) Spectra of the vertical and horizontal modes. (b) Their temporal shapes (top and bottom panels, respectively). The evolution of the vertical mode versus roundtrips is displayed in the temporal (c) and spectral (d) domains. The intracavity evolution of the VS over an entire cycle of the caterpillar motion: (e) in the vertical mode, and (f) in the horizontal one. The inset in Fig. 3(c) displays the intensity autocorrelation. The right panel of Fig. 3(d) displays the energy evolution versus roundtrips.

A fundamental question arises: why is the horizontal mode tightly localized under the action of strong LMC, unlike the extended damped pulse chain supported by weak LMC, cf. Figs. 3(b) and 2(d)? This can be explained by two points: (i) the energy exchange and mode mixing introduced by the strong LMC impart the confined profile of the vertical mode onto the horizontal one; (ii) the strong LMC, while enhancing the coupling between two modes, effectively suppresses the ITS-induced energy transfer within the horizontal mode. For instance [Fig. 3(f)], horizontal subpulse $P_2$ is strongly coupled to vertical $P_2$, being weakly coupled by ITS to the adjacent horizontal subpulse $P_3$.

*Experiments*.—The passive mode-locking ensured by a carbon nanotube is utilized to generate VSs in the laser. By adjusting LMC via PC, we first achieved heteronuclear VSs with dissimilar modes [Fig. 4(a)] (10-ps ITS, 75-mW pump power). The broadband vertical mode and comb-like horizontal one demonstrate consistency with the simulation. In addition to the dissimilar spectra of the two modes, the autocorrelations reveal that the vertical mode is a single pulse [Fig. 4(b)], whereas the horizontal mode exhibits multiple subpulses with a spacing of ~9.78 ps. The profile obtained from the frequency-resolved optical gating validates the existence of the damped pulse chain, which also confirms the spectral coherence of the horizontal mode (see further details in Sec. 2.2 in SM [62]). The synchronization between the dissimilar modes is evidenced by the identical repetition rate [Fig. 4(c)]. Interestingly, the interplay of LMC and ITS that induces the damped pulse chain holds significance for regulating the regular pulse packages in delayed-feedback semiconductor lasers [72].

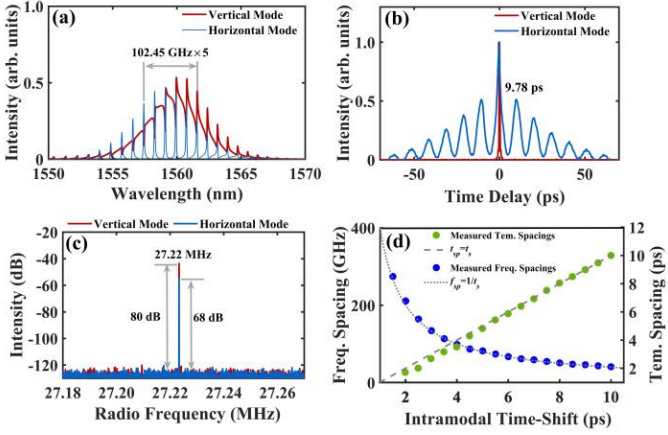
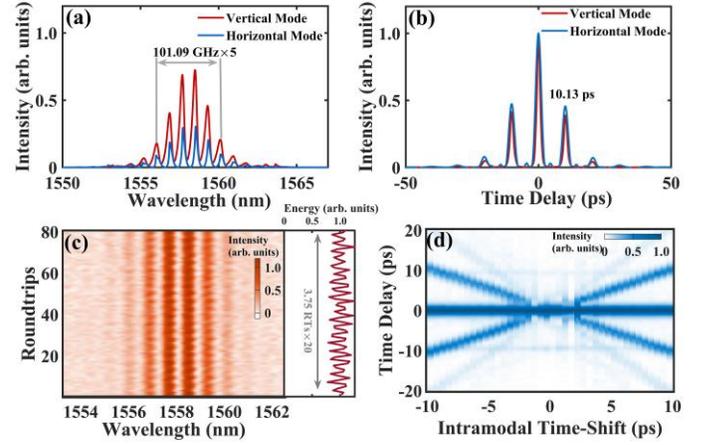

FIG. 4. Experimentally measured heteronuclear VS with the weak LMC at 75-mW pump power. (a) Spectra, (b) autocorrelation traces, (c) radio-frequency spectra corresponding to the 10-ps ITS. (d) Frequency and temporal spacings in the horizontal mode versus ITS.

FIG. 5. Experimentally measured homonuclear VS with the strong LMC at 75-mW pump power. (a) Optical spectra and (b) autocorrelation traces of two modes (10-ps ITS). (c) The real-time spectral evolution of the vertical mode, with the right panel showing the energy evolution. (d) Autocorrelation traces of the total pulse versus the values of ITS.

In the horizontal mode, the spectral and temporal spacings, $f_{sp}$ and $t_{sp}$, are measured as $f_{sp} = 1/t_s$ and $t_{sp} = t_s$ [Fig. 4(d)]. These relations are based on the phase-matching condition derived from the mean-field approximation, which is valid in the case of the weak LMC, considering the boundary condition of the resonator (see details in Sec. 2.3 of SM [68]):

$$\omega_j = \frac{2\pi j + \varphi_{Bi}}{t_s + D\omega_j/2}, \quad (2)$$

where $\omega_j$ is the resonant frequency for the horizontal mode, $\varphi_{Bi}$ is the birefringence phase shift, $D$ is the cavity dispersion, and $j$ is an integer. When $|t_s| \gg |D\omega_j/2| = 434$ fs, Eq. (2) yields $\omega_j \approx (2\pi j + \varphi_{Bi})/t_s$ with the uniform frequency spacing, $f_{sp} = 1/t_s$. Consequently, the temporal interferometric pattern has a spacing $t_{sp} = 1/f_{sp} = t_s$. Such a comb-like spectrum tunable via ITS can be further shaped to construct frequency lattices [73] and generate high-repetition-rate soliton crystals [74].

In agreement with the numerical prediction, enhancing the LMC strength via PC results in the formation of homonuclear VS, with similar profiles of the two polarization modes [Figs. 5(a) and 5(b)]. The autocorrelations indicate that the two modes are both soliton compounds localized in a smaller temporal region compared to the damped pulse chain in the weak-LMC-induced heteronuclear VS. The quasi-periodic spectral evolution of the vertical mode [Fig. 5(c)], also reflected by the energy evolution with a period of ~3.71 roundtrips [right panel of Fig. 5(c)], provides indirect evidence of the caterpillar motion. This pulsating VS also unlocks an alternative testbed for exploring breathers and chaos in nonlinear systems [75]. As predicted by simulations, the temporal structure of the homonuclear VS is deformable by tuning ITS confirmed by the intensity autocorrelations [Fig. 5(d)].

In experiments, an intermediate VS is also observed, exhibiting the horizontal mode with a comb-like spectrum coupled to the vertical mode with a modulated spectrum [Fig. 6(a)]. Compared to the homonuclear VS, the corresponding horizontal mode remains a damped pulse chain rather than the soliton compound [Fig. 6(b)]. Simulation results indicate that the LMC strength ranges from 0.05 to 0.17, see further details in Sec. 3 of SM [68]. Thereby, after the coupling energy to the vertical mode to form a multipulse compound (similar to the homonuclear case), the horizontal mode still has sufficient energy to ensure the ITS-induced damped pulse chain. Consequently, the intermediate VS exhibits critical behaviors between the heteronuclear and homonuclear VSs.

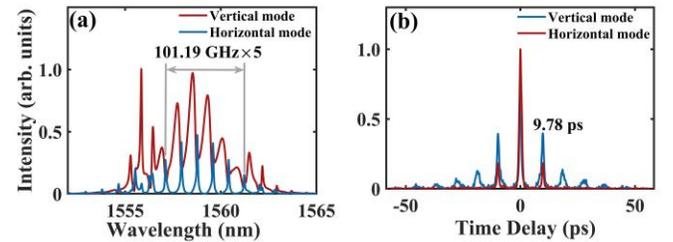

FIG. 6. Experimentally measured intermediate VS at 75-mW pump power and 10-ps ITS. (a) Optical spectra and (b) autocorrelation traces of two modes.

In conclusion, we have demonstrated the creation of VSs (vector solitons) in the laser, with the temporal and spectral structures that are deformable by the LMC (linear mode coupling) and ITS (intramodal time shift). The weak LMC in conjunction with ITS facilitates the heteronuclear VS, combining a narrow pulse in one polarization mode and a damped pulse chain in the other. The strong LMC+ITS results in the homonuclear VS built of similar soliton compounds exhibiting rolling-forward ("caterpillar") motions. Our work is of general interest in several aspects: (i) LMC+ITS equivalently creates periodic potential wells in both the temporal and frequency domains, and so holds promise for studying wave dynamics in photonic lattices [45,66], synthesizing soliton compounds [56], and tunable frequency combs [50]; (ii) Our VS laser is a non-Hermitian system with a tunable coupling strength, which has potential for uncovering new soliton patterns and phase transitions [76,77]; (iii) The mechanism underlying deformable VS can be transferred to other optical systems [2,26,27,78], and has a general guidance to soliton physics in Bose-Einstein condensates [7], plasma [79], and fluid dynamics [80].

## ACKNOWLEDGEMENTS

This work was supported by the National Natural Science Foundation of China (12274344, 12474339); Guangdong Basic and Applied Basic Research Foundation (2023A1515011517); Israel Science Foundation (grant No. 1695/22).

Contact author: *maodong@nwpu.edu.cn
Contact author: †jlzhao@nwpu.edu.cn

[1] J. Kang and G. Stegeman, Observation of Manakov Spatial Solitons in AlGaAs Planar Waveguides, Physical Review Letters **76**, 3699-3702 (1996).
[2] M. Marconi, J. Javaloyes, S. Barland, S. Balle, and M. Giudici, Vectorial dissipative solitons in vertical-cavity surface-emitting lasers with delays, Nature Photonics **9**, 450-455 (2015).
[3] K. Krupa, K. Nithyanandan, and P. Grelu, Vector dynamics of incoherent dissipative optical solitons, Optica **4**, 1239-1244 (2017).
[4] D. Mao, Z. Yuan, K. Dai, Y. Chen, H. Ma, Q. Ling, J. Zheng, Y. Zhang, D. Chen, Y. Cui, Z. Sun, and B. Malomed, Temporal and spatiotemporal soliton molecules in ultrafast fibre lasers, Nanophotonics **14**, 677-706 (2025).
[5] B. Deng, J. Raney, V. Tournat, and K. Bertoldi, Elastic Vector Solitons in Soft Architected Materials, Physical Review Letters **118**, 204102 (2017).
[6] B. Deng, V. Tournat, P. Wang, and K. Bertoldi, Anomalous Collisions of Elastic Vector Solitons in Mechanical Metamaterials, Physical Review Letters **122**, 044101 (2019).
[7] X. Zhang, X. Hu, X. Liu, and W. Liu, Vector solitons in two-component Bose-Einstein condensates with tunable interactions and harmonic potential, Physical Review A **79**, 033630 (2009).
[8] S. Lannig, C. Schmied, M. Prufer, P. Kunkel, R. Strohmaier, H. Strobel, T. Gasenzer, P. Kevrekidis, and M. Oberthaler, Collisions of Three-Component Vector Solitons in Bose-Einstein Condensates, Physical Review Letters **125**, 170401 (2020).
[9] B. Malomed and S. Wabnitz, Soliton annihilation and fusion from resonant inelastic collisions in birefringent optical fibers, Optics Letters **16**, 1388-1390 (1991).
[10] J. Yang and Y. Tan, Fractal Structure in the Collision of Vector Solitons, Physical Review Letters **85**, 3624-3627 (2000).
[11] C. Anastassiou, M. Segev, K. Steiglitz, J. Giordmaine, M. Mitchell, M. Shih, S. Lan, and J. Martin, Energy-Exchange Interactions between Colliding Vector Solitons, Physical Review Letters **83**, 2332-2335 (1999).
[12] C. Anastassiou, J. Fleischer, T. Carmon, M. Segev, and K. Steiglitz, Information transfer via cascaded collisions of vector solitons, Optics Letters **26**, 1498-1500 (2001).
[13] D. Rand, I. Glesk, C. Bres, D. Nolan, X. Chen, J. Koh, J. Fleischer, K. Steiglitz, and P. Prucnal, Observation of temporal vector soliton propagation and collision in birefringent fiber, Physical Review Letters **98**, 053902 (2007).
[14] M. Mitchell, M. Segev, and D. Christodoulides, Observation of Multihump Multimode Solitons, Physical Review Letters **80**, 4657-4660 (1998).
[15] A. Desyatnikov and Y. Kivshar, Necklace-ring vector solitons, Physical Review Letters **87**, 033901 (2001).
[16] E. Ostrovskaya, Y. Kivshar, D. Skryabin, and W. Firth, Stability of Multihump Optical Solitons, Physical Review Letters **83**, 296-299 (1999).
[17] W. Krolikowski, E. Ostrovskaya, C. Weilnau, M. Geisser, G. McCarthy, Y. Kivshar, C. Denz, and B. Luther-Davies, Observation of Dipole-Mode Vector Solitons, Physical Review Letters **85**, 1424-1427 (2000).
[18] C. Cambournac, T. Sylvestre, H. Maillotte, B. Vanderlinden, P. Kockaert, P. Emplit, and M. Haelterman, Symmetry-breaking instability of multimode vector solitons, Physical Review Letters **89**, 083901 (2002).
[19] M. Shalaby and A. Barthelemy, Observation of the Self-Guided Propagation of a Dark and Bright Spatial Soliton Pair in a Focusing Nonlinear Medium, IEEE Journal of Quantum Electronics **28**, 2736-2741 (1992).
[20] Y. Kivshar and S. Turitsyn, Vector dark solitons, Optics Letters **18**, 337-339 (1993).
[21] C. Menyuk, Stability of solitons in birefringent optical fibers. I: Equal propagation amplitudes, Optics Letters **12**, 614-616 (1987).
[22] C. Menyuk, Stability of solitons in birefringent optical fibers. II. Arbitrary amplitudes, Journal of Optical Society of America B **5**, 392-402 (1988).
[23] M. Tratnik and J. Sipe, Bound solitary waves in a birefringent optical fiber, Physical Review A **38**, 2011-2017 (1988).
[24] D. Mihalache, D. Mazilu, and L. Torner, Stability of Walking Vector Solitons, Physical Review Letters **81**, 4353-4356 (1998).
[25] S. Cundiff, B. Collings, N. Akhmediev, J. Soto-Crespo, K. Bergman, and W. Knox, Observation of Polarization-Locked Vector Solitons in an Optical Fiber, Physical Review Letters **82**, 3988-3991 (1999).
[26] G. Xu, A. U. Nielsen, B. Garbin, L. Hill, G. L. Oppo, J. Fatome, S. G. Murdoch, S. Coen, and M. Erkintalo, Spontaneous symmetry breaking of dissipative optical solitons in a two-component Kerr resonator, Nature Communications **12**, 4023 (2021).
[27] F. Copie, M. Woodley, L. Del Bino, J. Silver, S. Zhang, and P. Del'Haye, Interplay of Polarization and Time-Reversal Symmetry Breaking in Synchronously Pumped Ring Resonators, Physical Review Letters **122**, 013905 (2019).
[28] J. Haus, G. Shaulov, E. Kuzin, and J. Sanchez-Mondragon, Vector soliton fiber lasers, Optics Letters **24**, 376-378 (1999).
[29] N. Akhmediev, J. Soto-Crespo, S. Cundiff, B. Collings, and W. Knox, Phase locking and periodic evolution of solitons in passively mode-locked fiber lasers with a semiconductor saturable absorber, Optics Letters **23**, 852-854 (1998).
[30] D. Tang, H. Zhang, L. Zhao, and X. Wu, Observation of high-order polarization-locked vector solitons in a fiber laser, Physical Review Letters **101**, 153904 (2008).
[31] L. Zhao, D. Tang, X. Wu, H. Zhang, and H. Tam, Coexistence of

polarization-locked and polarization-rotating vector solitons in a fiber laser with SESAM, Optics Letters **34**, 3059-3061 (2009).
[32] L. Zhao, D. Tang, H. Zhang, X. Wu, and N. Xiang, Soliton trapping in fiber lasers, Optics Express **16**, 9528-9533 (2008).
[33] M. Liu, A. Luo, Z Luo, and W. Xu, Dynamic trapping of a polarization rotation vector soliton in a fiber laser, Optics Letters **42**, 330-333 (2017).
[34] S. Sergeyev, C. Mou, E. Turitsyna, A. Rozhin, S. Turitsyn, and K. Blow, Spiral attractor created by vector solitons, Light: Science & Applications **3**, e131 (2014).
[35] Z. Huang, S. Sergeyev, Q. Wang, H. Kbashi, D. Stoliarov, Q. Huang, Y. Dai, Z. Yan, and C. Mou, Dissipative soliton breathing dynamics driven by desynchronization of orthogonal polarization states, Advanced Photonics Nexus **2**, 066007 (2023).
[36] S. Zhang, T. Bi, G. Ghalanos, N. Moroney, L. Del Bino, and P. Del'Haye, Dark-Bright Soliton Bound States in a Microresonator, Physical Review Letters **128**, 033901 (2022).
[37] X. Hu, J. Guo, J. Wang, J. Ma, L. Zhao, S. Yoo, and D. Tang, Novel optical soliton molecules formed in a fiber laser with near-zero net cavity dispersion, Light: Science & Applications **12**, 38 (2023).
[38] G. Shao, Y. Song, J. Guo, L. Zhao, D. Shen, and D. Tang, Induced dark solitary pulse in an anomalous dispersion cavity fiber laser, Optics Express **23**, 28430-28437 (2015).
[39] D. Mao, Z. He, Y. Zhang, Y. Du, C. Zeng, L. Yun, Z. Luo, T. Li, Z. Sun, and J. Zhao, Phase-matching-induced near-chirp-free solitons in normal-dispersion fiber lasers, Light: Science & Applications **11**, 25 (2022).
[40] S. Sergeyev, H. Kbashi, N. Tarasov, Y. Loiko, and S. Kolpakov, Vector-Resonance-Multimode Instability, Physical Review Letters **118**, 033904 (2017).
[41] R. Parr and W. Yang, Density-functional theory of the electronic structure of molecules, Annual Review of Physical Chemistry **46**, 701-728 (1995).
[42] C. Stehle, C. Zimmermann, and S. Slama, Cooperative coupling of ultracold atoms and surface plasmons, Nature Physics **10**, 937-942 (2014).
[43] F. Arute, F. Arya, K. Babbush *et al.*, Quantum supremacy using a programmable superconducting processor, Nature **574**, 505-510 (2019).
[44] P. Smith, Mode-Locking of Lasers, Proceedings of the IEEE **58**, 1342-1363 (1970).
[45] A. Senanian, L. Wright, P. Wade, H. Doyle, and P. McMahon, Programmable large-scale simulation of bosonic transport in optical synthetic frequency lattices, Nature Physics **19**, 1333-1339 (2023).
[46] A. Regensburger, C. Bersch, M. Miri, G. Onishchukov, D. Christodoulides, and U. Peschel, Parity-time synthetic photonic lattices, Nature **488**, 167-171 (2012).
[47] C. Leefmans, M. Parto, J. Williams, G. Li, A. Dutt, F. Nori, and A. Marandi, Topological temporally mode-locked laser, Nature Physics (2024).
[48] T. Erdogan, Fiber Grating Spectra, IEEE Journal of Lightwave Technology **15**, 1277-1294 (1997).
[49] A. Tikan, J. Riemensberger, K. Komagata, S. Hönl, M. Churaev, C. Skehan, H. Guo, R. Wang, J. Liu, P. Seidler, and T. Kippenberg, Emergent nonlinear phenomena in a driven dissipative photonic dimer, Nature Physics **17**, 604-610 (2021).
[50] T. Letsou, D. Kazakov, P. Ratra, L. Columbo, M. Brambilla, F. Prati, C. Rimoldi, N. Opačak, H. Everitt, M. Piccardo, B. Schwarz, and F. Capasso, Hybridized Soliton Lasing in Coupled Semiconductor Lasers, Physical Review Letters **134**, 023802 (2025).
[51] L. Wright, P. Sidorenko, H. Pourbeyram, Z. Ziegler, A. Isichenko, B. Malomed, C. Menyuk, D. Christodoulides, and F. Wise, Mechanisms of spatiotemporal mode-locking, Nature Physics **16**, 565-570 (2020).
[52] L. Liu, Z. He, Q. Gao, Y. Du, C. Zeng, and D. Mao, Self-consistent soliton evolution in single-two-mode fiber lasers, Optics Letters **47**, 6369-6372 (2022).
[53] C. Gao, B. Cao, Y. Ding, X. Xiao, D. Yang, H. Fei, C. Yang, and C. Bao, All-step-index-fiber spatiotemporally mode-locked laser, Optica **10**, 356-363 (2023).
[54] Y. Du, Z. He, Q. Gao, H. Zhang, C. Zeng, D. Mao, and J. Zhao, Emergent Phenomena of Vector Solitons Induced by the Linear Coupling, Laser & Photonics Reviews **17**, 2300076 (2023).
[55] M. Marconi, J. Javaloyes, S. Balle, and M. Giudici, How lasing localized structures evolve out of passive mode locking, Physical Review Letters **112**, 223901 (2014).
[56] M. Pang, W. He, X. Jiang, and P. Russell, All-optical bit storage in a fibre laser by optomechanically bound states of solitons, Nature Photonics **10**, 454-458 (2016).
[57] J. Javaloyes, M. Marconi, and M. Giudici, Nonlocality Induces Chains of Nested Dissipative Solitons, Physical Review Letters **119**, 033904 (2017).
[58] L. Nimmesgern, C. Beckh, H. Kempf, A. Leitenstorfer, and G. Herink, Soliton molecules in femtosecond fiber lasers: universal binding mechanism and direct electronic control, Optica **8**, 1334-1339 (2021).
[59] A. Schreiber, K. Cassemiro, V. Potocek, A. Gabris, P. Mosley, E. Andersson, I. Jex, and C. Silberhorn, Photons walking the line: a quantum walk with adjustable coin operations, Physical Review Letters **104**, 050502 (2010).
[60] W. Weng, R. Bouchand, E. Lucas, E. Obrzud, T. Herr, and T. Kippenberg, Heteronuclear soliton molecules in optical microresonators, Nature Communications **11**, 2402 (2020).
[61] F. Kurtz, C. Ropers, and G. Herink, Resonant excitation and all-optical switching of femtosecond soliton molecules, Nature Photonics **14**, 9-13 (2019).
[62] S. Evangelides, L. Mollenauer, J. Gordon, and N. Bergano, Polarization Multiplexing with Solitons, IEEE Journal of Lightwave Technology **10**, 28-35 (1992).
[63] M. Gilles, P. Bony, J. Garnier, A. Picozzi, M. Guasoni, and J. Fatome, Polarization domain walls in optical fibres as topological bits for data transmission, Nature Photonics **11**, 102-107 (2017).
[64] M. Kowalczyk, Ł. Sterczewski, X. Zhang, V. Petrov, Z. Wang, and J. Sotor, Dual-Comb Femtosecond Solid-State Laser with Inherent Polarization-Multiplexing, Laser & Photonics Reviews **15**, 2000441 (2021).
[65] G. Pu, L. Yi, L. Zhang, and W. Hu, Intelligent programmable mode-locked fiber laser with a human-like algorithm, Optica **6**, 362-369 (2019).
[66] S. Wang, B. Wang, C. Liu, C. Qin, L. Zhao, W. Liu, S. Longhi, and P. Lu, Nonlinear Non-Hermitian Skin Effect and Skin Solitons in Temporal Photonic Feedforward Lattices, Physical Review Letters **134**, 243805 (2025).
[67] D. Cheng, K. Wang, C. Roques-Carmes, E. Lustig, O. Long, H. Wang, and S. Fan, Non-Abelian lattice gauge fields in photonic synthetic frequency dimensions, Nature **637**, 52-56 (2025).
[68] See Supplemental Material at http://link.aps.org/supplemental/xxx for more details on experimental setups, measured results and numerical simulations of the vector-soliton under the linear mode-couping, which includes Refs. [54, 69].
[69] A. Hasegawa and Y. Kodama, Guiding-center soliton, Physical Review Letters **66**, 161-164 (1991).
[70] S. Hamdi, A. Coillet, B. Cluzel, P. Grelu, and P. Colman, Superlocalization Reveals Long-Range Synchronization of Vibrating Soliton Molecules, Physical Review Letters **128**, 213902 (2022).
[71] K. Krupa, K. Nithyanandan, U. Andral, P. Tchofo-Dinda, and P. Grelu, Real-Time Observation of Internal Motion within Ultrafast Dissipative Optical Soliton Molecules, Physical Review Letters **118**, 243901 (2017).

[72] T. Heil, I. Fischer, W. Elsäßer, and A. Gavrielides, Dynamics of Semiconductor Lasers Subject to Delayed Optical Feedback: The Short Cavity Regime, Physical Review Letters **87**, 243901 (2001).

[73] A. Balčytis, T. Ozawa, Y. Ota, S. Iwamoto, J. Maeda, and T. Baba, Synthetic dimension band structures on a Si CMOS photonic platform, Science Advances **8**, 1-9 (2022).

[74] D. Cole, E. Lamb, P. Del'Haye, S. Diddams, and S. Papp, Soliton crystals in Kerr resonators, Nature Photonics **11**, 671-676 (2017).

[75] H. Kang, A. Zhou, Y. Zhang, X. Wu, B. Yuan, J. Peng, C. Finot, S. Boscolo, and H. Zeng, Observation of Optical Chaotic Solitons and Modulated Subharmonic Route to Chaos in Mode-Locked Laser, Physical Review Letters **133**, 263801 (2024).

[76] L. Feng, R. El-Ganainy, and L. Ge, Non-Hermitian photonics based on parity–time symmetry, Nature Photonics **11**, 752-762 (2017).

[77] L. Li, Y. Cao, Y. Zhi, J. Zhang, Y. Zou, X. Feng, B. Guan, and J. Yao, Polarimetric parity-time symmetry in a photonic system, Light: Science & Applications **9**, 169 (2020).

[78] H. Haig, P. Sidorenko, A. Dhar, N. Choudhury, R. Sen, D. Christodoulides, and F. Wise, Multimode Mamyshev oscillator, Optics Letters **47**, 46-49 (2022).

[79] V. Berezhiani, Z. Osmanov, S. Mahajan, and S. Mikeladze, Solitary structure formation and self-guiding of electromagnetic beam in highly degenerate electron plasma, Physics of Plasmas **28**, 052104 (2021).

[80] R. Barros, W. Choi, and P. Milewski, Strongly nonlinear effects on internal solitary waves in three-layer flows, Journal of Fluid Mechanics **883**, A16 (2019).